\documentclass[aps,prl,10pt,amsmath,amssymb,floatfix,twocolumn,
superscriptaddress]{revtex4}

\usepackage[english]{babel}
\usepackage[utf8x]{inputenc}
\usepackage{graphicx}
\usepackage{amsmath}
\usepackage{amssymb}
\usepackage{nicefrac}
\usepackage{hyperref}
\usepackage{xspace}

\usepackage[uppercase,tiny]{titlesec}
\titlespacing*{\section}{0pt}{12pt}{4pt}

\let\AAt\AA
\renewcommand{\AA}{\ifmmode{\mathrm{\AAt}}\else{\AAt}\fi}
\newcommand{\var}[1]{\ensuremath{#1}\xspace}
\newcommand{\unit}[1]{\ensuremath{\mathrm{#1}}\xspace}
\newcommand{\half}{\ensuremath{\nicefrac{1}{2}}\xspace}
\newcommand{\chem}[1]{\ensuremath{\rm #1}\xspace}
\newcommand{\mB}{{\rm \mu_B}}
\newcommand{\mS}{{\rm \mu S}}
\newcommand{\hBN}{{\it h}\text{--}BN}

\preprint{Spin Switch v2.1}

\begin{document}

	\title{Potential Energy Driven Spin Manipulation via a Controllable Hydrogen Ligand}
	
	\author{Peter Jacobson}
	\email[Correspondence to Peter Jacobson ]{peter.jacobson@uni-graz.at}
	\email[and Markus Ternes ]{m.ternes@fkf.mpg.de}
	\altaffiliation[Current affiliation: ]{Institute of Chemistry, University of Graz, Graz, Austria}
	\thanks{P.J. and M.M. contributed equally}
	\affiliation{Max Planck Institute for Solid State Research, Stuttgart, Germany}
	\author{Matthias Muenks}
	\affiliation{Max Planck Institute for Solid State Research, Stuttgart, Germany}
	\author{Gennadii Laskin}
	\affiliation{Max Planck Institute for Solid State Research, Stuttgart, Germany}
	\author{Oleg~O.~Brovko}
	\affiliation{The Abdus Salam International Centre for Theoretical Physics (ICTP), Trieste, Italy}
	\author{Valeri~S.~Stepanyuk}
	\affiliation{Max Planck Institute of Microstructure Physics, Halle, Germany}
	\author{Markus Ternes}
	\affiliation{Max Planck Institute for Solid State Research, Stuttgart, Germany}
	\author{Klaus Kern}
	\affiliation{Max Planck Institute for Solid State Research, Stuttgart, Germany}
	\affiliation{Institute de Physique, \'Ecole Polytechnique F\'ed\'erale de Lausanne, Lausanne, Switzerland}
	
 \begin{abstract}
	Spin-bearing molecules can be stabilized on surfaces and in junctions with desirable properties such as a net spin that can be adjusted by external stimuli.  Using scanning probes, initial and final spin states can be deduced from topographic or spectroscopic data, but how the system transitioned between these states is largely unknown.  Here we address this question by manipulating the total spin of magnetic cobalt hydride complexes on a corrugated boron nitride surface with a hydrogen-functionalized scanning probe tip by simultaneously tracking force and conductance. When the additional hydrogen ligand is brought close to the cobalt monohydride, switching between a correlated $\var{S} = \half$ Kondo state, where host electrons screen the magnetic moment, and a $\var{S} = 1$ state with magnetocrystalline anisotropy is observed.  We show that the total spin changes when the system is transferred onto a new potential energy surface defined by the position of the hydrogen in the junction.  These results show how and why chemically functionalized tips are an effective tool to manipulate adatoms and molecules, and a promising new method to selectively tune spin systems.
 \end{abstract}

	\maketitle
	
\section*{Introduction}
	
	The magnetic behavior of adatoms and single molecular magnets on surfaces is usually defined by static parameters such as the local symmetry, the spin-orbit interaction, or the exchange coupling with the electron bath of the host.~\cite{Otte2008,Kahle2011,Rau2014,Oberg2014,Jacobson2015} However, there is widespread interest in actively controlling molecular and adatom spin states for switching applications.~\cite{Heinrich2015,Parks2010} Beyond imaging and spectroscopy, scanning probes are atomically precise manipulation tools.~\cite{Eigler1990,Eigler1991} When manipulation and spectroscopy operate in tandem, it is possible to observe the formation of chemical bonds and continuously tune the exchange interaction between magnetic impurities.~\cite{Custance2009,Bork2011,Choi2012,Muenks2016} Tip functionalization, now routinely used to create chemically precise contacts where a molecule acts as the transducer, is one promising method for controlling spins.~\cite{Wagner2015,Bartels1997,Weymouth2014,Kichin2011,Weiss2010} This strategy has its roots in small molecule adsorption on metal bearing porphyrins and phthalocyanines~\cite{Wende2007,Wackerlin2012} and capitalizes on two strengths of local probes: the ability to address specific atomic sites and the variable width of the tunnel junction. With magnetic adatoms gaining prominence as model quantum systems, it is highly desirable to understand how chemically reactive probes couple to and influence the measurement process and eventually control the resulting magnetic state. 

	Here, we reversibly control the total spin of cobalt hydride (\chem{CoH}) spin centers adsorbed on the \chem{\hBN/Rh(111)} moir\'e by manipulating a single hydrogen atom with the tip of a combined scanning tunneling (STM) and non-contact atomic force microscope (AFM). As the distance $z$ between the probing tip and the \chem{CoH} complex decreases, hydrogen initially adsorbed on the tip apex weakly bonds to the \chem{CoH} complex, inducing rapid transitions between a correlated $\var{S} = \half$ Kondo state and an anisotropic $\var{S} = 1$ state. Local spectroscopy identifies a stable total spin at high and low values of the conductance, while at intermediate conductance dynamic switching is observed. Combining conductance-distance, \var{G(z)}, and force-distance, $F(z)$, measurements together with density functional theory (DFT) calculations, we unravel the microscopic potential energy landscapes present within the tunnel junction. We demonstrate that by coupling a functionalized tip to an undercoordinated adatom, the reactivity of the adatom can be harnessed to drive transitions between different total spin states.  The spin within the tunnel junction can therefore be actively monitored and reversibly controlled with single atom precision.

\section*{Results}
	
	\begin{figure*}
		\center{\includegraphics{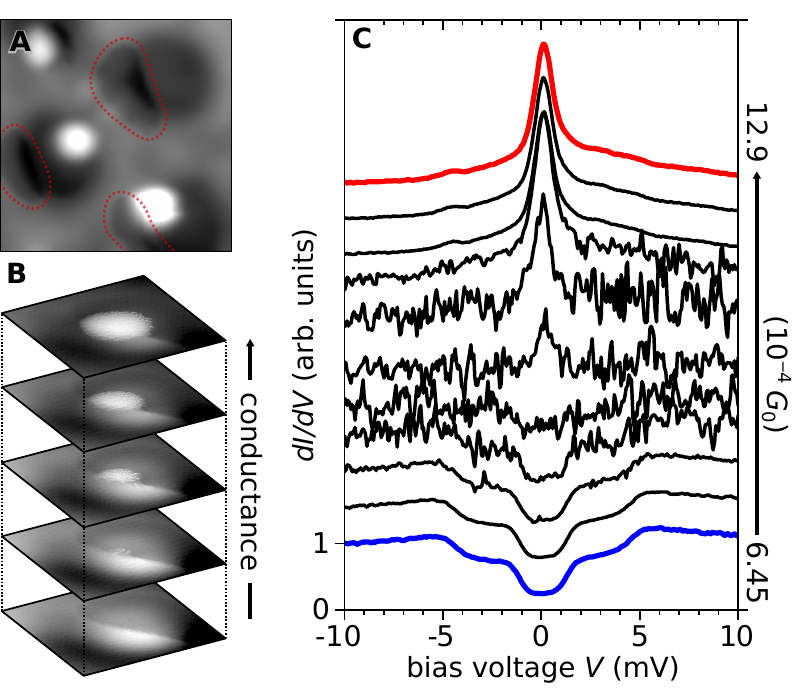}}
		\caption{\textbf{Influence of Hydrogen Functionalized Tips on Imaging and Spectroscopy.} (A) Constant current STM image (approximately $5 \times 5~\unit{nm^2}$; 	$\var{V} = -15~\unit{mV}$, $\var{I} = 20~\unit{pA}$; $\var{G} = 1.72 \times 10^{-5}~\var{G_0}$) of \chem{CoH} complexes on the \chem{\hBN/Rh(111)} moir\'e obtained with a hydrogen functionalized tip. Areas with enhanced contrast due to hydrogen in the junction are outlined in red. (B) Constant current STM images ($1.2 \times 1.2~\unit{nm^2}$, top to bottom: $\var{V} = -0.3$, $-0.7$, $-1.0$, $-1.3$, $-1.6~\unit{mV}$, $\var{I} = 20~\unit{pA}$, corresponding to $\var{G} = 8.60$, $3.69$, $2.58$, $1.99$, $1.61 \times 10^{-4}~\var{G_0}$) of a \chem{CoH} complex highlighting the strong conductance (tip-sample distance) dependence of imaging with a hydrogen-functionalized tip. (C) Local spectroscopy obtained on the \chem{CoH} complex in panel (B), the tip was centered on the dark lobe ($\var{G} = 1.61 \times 10^{-4}~\var{G_0}$). At $\var{G} = 6.45 \times 10^{-4}~\var{G_0}$ (blue), a set of double steps is observed, indicative of a spin 1 complex with magnetic anisotropy. Increasing the conductance in steps of $\var{\Delta G} = 0.16 \times 10^{-4}~\var{G_0}$ leads to the unstable spectroscopy until a spin $\half$ Kondo peak emerges at high conductance (red, $\var{G} = 12.9 \times 10^{-4}~\var{G_0}$). All spectra are normalized to the differential conductance at $-10~\unit{mV}$, normalized spectra are offset by $0.5$.}
		\label{fig:01}
	\end{figure*}

	Figure~\ref{fig:01}A shows a constant current image of \chem{CoH} complexes on \chem{\hBN/Rh(111)}. The lattice mismatch between the \chem{Rh(111)} substrate and the single monolayer of \chem{\hBN} results in a strongly corrugated surface with $3.2~\unit{nm}$ periodicity on which the \chem{CoH} complexes appear as bright protrusions. A clear indication of hydrogen adsorption on the tip apex are the sharp changes in tip height, reduced by $20~\mathrm{pm}$ (Figure~\ref{fig:01}A, red dashes), while imaging the \chem{\hBN/Rh(111)} surface in constant current mode.~\cite{Temirov2008} Figure~\ref{fig:01}B shows an individual \chem{CoH} complex located near the rim-valley boundary of \chem{\hBN/Rh(111)} imaged with a hydrogen-functionalized tip. At low junction conductance ($\var{G} = \var{I_S/V_S} = 1.61 \times 10^{-4}~\var{G_0}$; $\var{G_0} = 77.48~\unit{\mS}$, the quantum of conductance), corresponding to relatively large tip-sample separations \var{z}, the increased contrast due to hydrogen in the junction partially overlaps a \chem{CoH} (Figure~\ref{fig:01}B, bottom panel). As \var{G} is increased and $z$ decreases, this boundary region transitions to a noise speckled circle with a brighter appearance, \textit{i.e.} larger $z$-height, to compensate for an overall increase in the conductance. Given the strong \var{G} dependence within such a narrow range, these results hint that the observed contrast is not solely due to the local topography, but also due to mechanical and electronic changes in the junction. Indeed, these images are qualitatively similar to measurements of undercoordinated metal adatoms in the presence of adsorbed hydrogen.~\cite{Pivetta2007,Esat2015} As the hydrogen content of the \chem{CoH_x} complex governs the spin state,~\cite{Jacobson2015} \var(dI/dV) spectroscopy was performed while varying the setpoint conductance \var{G} with the tip positioned over the central region. At the lowest conductance, $\var{G} = 6.45 \times 10^{-4}~\var{G_0}$ (Figure~\ref{fig:01}C, bottom curve), the spectra show two symmetric steps around zero bias with increasing differential conductance. These steps originate from the inelastic spin excitations of a \chem{CoH} complex with total spin $\var{S} = 1$ where magnetocrystalline anisotropy has removed the $3d$ level degeneracy. Increasing \var{G} results in progressively unstable spectra until the emergence of a stable zero bias peak at $\var{G} = 12.9 \times 10^{-4}~\var{G_0}$, identified as a $\var{S} = \half$ Kondo resonance of \chem{CoH_2}. ~\cite{Jacobson2015} This transition is fully reversible and the initial $\var{S} = 1$ total spin state restored when the junction conductance is reduced (see Fig.~\ref{fig:S01}). We observe a metastable state, when $\var{G}$ is between $8 \times 10^{-4}~\var{G_0}$ and $11 \times 10^{-4}~\var{G_0}$, where the hydride complex randomly transitions between the $\var{S} = 1$ and $\var{S} = \half$ states on a timescale of $100~\unit{ms}$. The change in tip-sample separation for this conductance range corresponds to a $\Delta z$ of less than $25~\unit{pm}$. Note, that this metastable behavior does not depend on the bias voltage during the spectroscopic measurement. Differential conductance (\var{dI/dV}) spectroscopy not only identifies the spin state, but it also aids in the interpretation of the STM images in Figure~\ref{fig:01}B. The constant current images in Figure~\ref{fig:01}B were obtained over a bias range ($0.3-1.6~\unit{mV}$) where the topographic appearance is closely linked to features in the \var{dI/dV} measurements and therefore, at small bias voltages, is dominated by the Kondo resonance.

	\begin{figure*}
		\center{\includegraphics{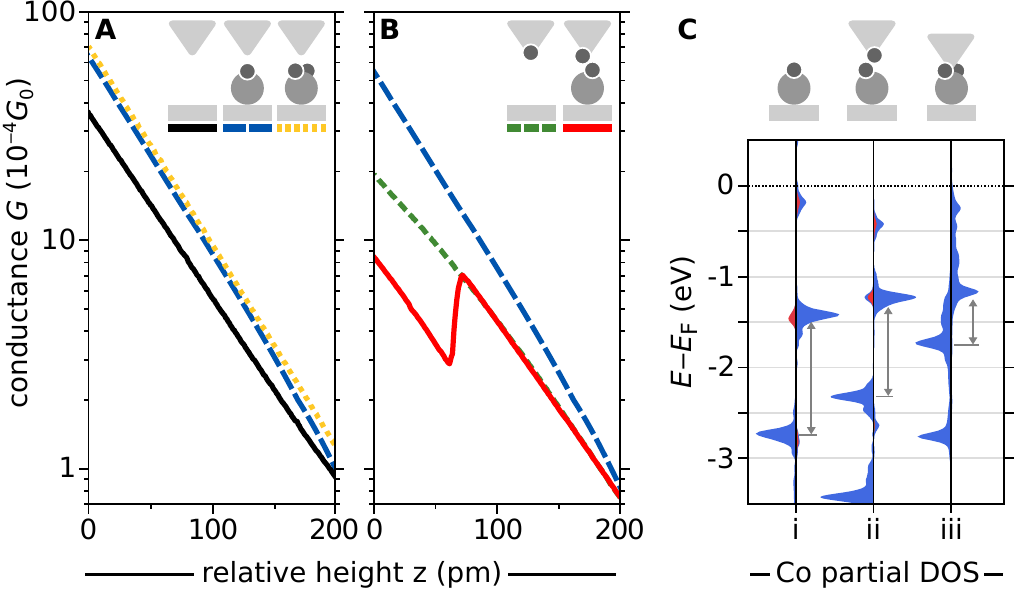}}
		\caption{\textbf{Conductance -- distance spectroscopy.} (A) Conductance-distance, \var{G(z)}, curves obtained with a bare \chem{Pt} tip on \chem{\hBN} (black), \chem{CoH} (dashed blue), and \chem{CoH_2} (dotted yellow) at a tip-sample bias $\var{V}= -10~\unit{mV}$. (B) Using a functionalized tip, \chem{CoH+H_{tip}} (red), a conductance discontinuity, corresponding to the $\var{S} = 1$ to $\var{S} = \half$ total spin change, is observed at a relative height \var{z} of $70~\unit{pm}$. The functionalized tip approaching the substrate, \chem{\hBN+H_{tip}} (dashed green), shows no discontinuity and a non-exponential character. For direct comparison, the \chem{CoH} \var{G(z)} measurement from panel (A) is plotted again (dashed blue). Inverse decay constants, $\var{\kappa_G}~\unit{[nm^{-1}]}$: (A) \chem{\hBN} (black): $8.7 \pm 0.1$; \chem{CoH} (dashed blue): $9.9 \pm 0.1$; \chem{CoH_2} (dotted yellow): $9.8 \pm 0.1$; (B) \chem{\hBN+H_{tip}} (dashed green): $6.6 \pm 0.3$ ($0 < \var{z} < 70~\unit{pm}$), $7.7 \pm 0.4$ ($70 < \var{z} < 200~\unit{pm}$); \chem{CoH+H_{tip}} (red): $6.9 \pm 0.4$ ($0 < z < 70~\unit{pm}$), $8.5 \pm 0.5$ ($70 < \var{z} < 200~\unit{pm}$). The color- and linetype-coded insets schematically depict the junction geometry. (C) Plots of the majority (left) and minority (right) spin projected density of states (blue: $d$ orbitals, red: $sp$ orbitals) of a $\var{S} = 1$ \chem{CoH} complex, magnetic moment of $2.0~\mB$,  without tip (i), one with a slight magnetic moment reduction ($1.6~\mB$) due to the presence of a hydrogen functionalized tip (ii), transition from $\var{S} = 1$ to $\var{S} = \half$ ($1.2~\mB$) at close tip distances (iii). The change in Stoner splitting between majority and minority bands is schematically depicted with vertical grey arrows.}
		\label{fig:02}
	\end{figure*}

	To investigate the switching behavior in detail, \var{G(z)} measurements were performed over \chem{CoH_x} complexes and bare \chem{\hBN}. Approaching \chem{\hBN} as well as \chem{CoH_x} complexes with a bare tip reveals a strictly exponential increase in conductance, $G(z) = G_0 \exp(-2\kappa_G\, (z_0 + z))$ with $\kappa_G$ the decay rate, and \var{z_0} the tip-height at the initial setpoint conductance \var{G} (Figure~\ref{fig:02}A). Functionalizing the tip apex with hydrogen alters the junction conductance characteristics, with \var{G(z)} showing a less than exponential increase and a reduced \var{\kappa_G} compared to data obtained with a bare tip (Figure~\ref{fig:02}B). This characteristic behavior is similar to the observations of Weiss \textit{et al.} on a complex organic molecule with a hydrogen-functionalized tip.~\cite{Weiss2010} However, approaching a \chem{CoH} complex with a hydrogen-functionalized tip, \var{G(z)} closely follows the \chem{\hBN} trace until the conductance rapidly decreases by a factor of $2.5$, indicating the $\var{S} = 1$ to $\var{S} = \half$ transition (Figure~\ref{fig:02}B, red) observed by local spectroscopy in Figure~\ref{fig:01}C.  Therefore, the drop in conductance stems from both the direct transfer of hydrogen within the junction, altering the geometry and modifying the tunnel barrier, and the relative change in total conductance between the \chem{CoH} and \chem{CoH_2} complexes.

	To understand the electronic structure modification in the spin switching process, we performed density functional theory calculations for various representative junction geometries (Figure~\ref{fig:02}C). A \chem{CoH} complex on \chem{\hBN} exhibits a nearly free-atom-like electronic structure (Figure~\ref{fig:02}C-i) with Stoner split $3d$ levels giving it a magnetic moment of $2.0~\mB$ (Bohr magneton). ~\cite{Jacobson2015} Approaching the complex with a Pt tip which has a hydrogen atom bound to its apex (Figure~\ref{fig:02}C-ii) gradually introduces indirect bonding via the \chem{H} of the \chem{CoH} complex, resulting in the reduction of the \chem{Co} magnetic moment (Figure~\ref{fig:02}C-ii). The magnetic moment of Co remains close to $2~\mB$, an effective $\var{S} = 1$ state, until the approaching functionalized tip brings the \chem{H} $sp$ orbitals into direct overlap with the \chem{Co} $d$ orbitals (Figure~\ref{fig:02}C-iii). Bonding of two \chem{H} atoms to \chem{Co}, as shown in our previous work, is strong enough to partially quench the magnetic moment of \chem{Co}, reducing it to $1.2~\mB$, an effective $\var{S} = \half$ state.

	\begin{figure*}
		\center{\includegraphics{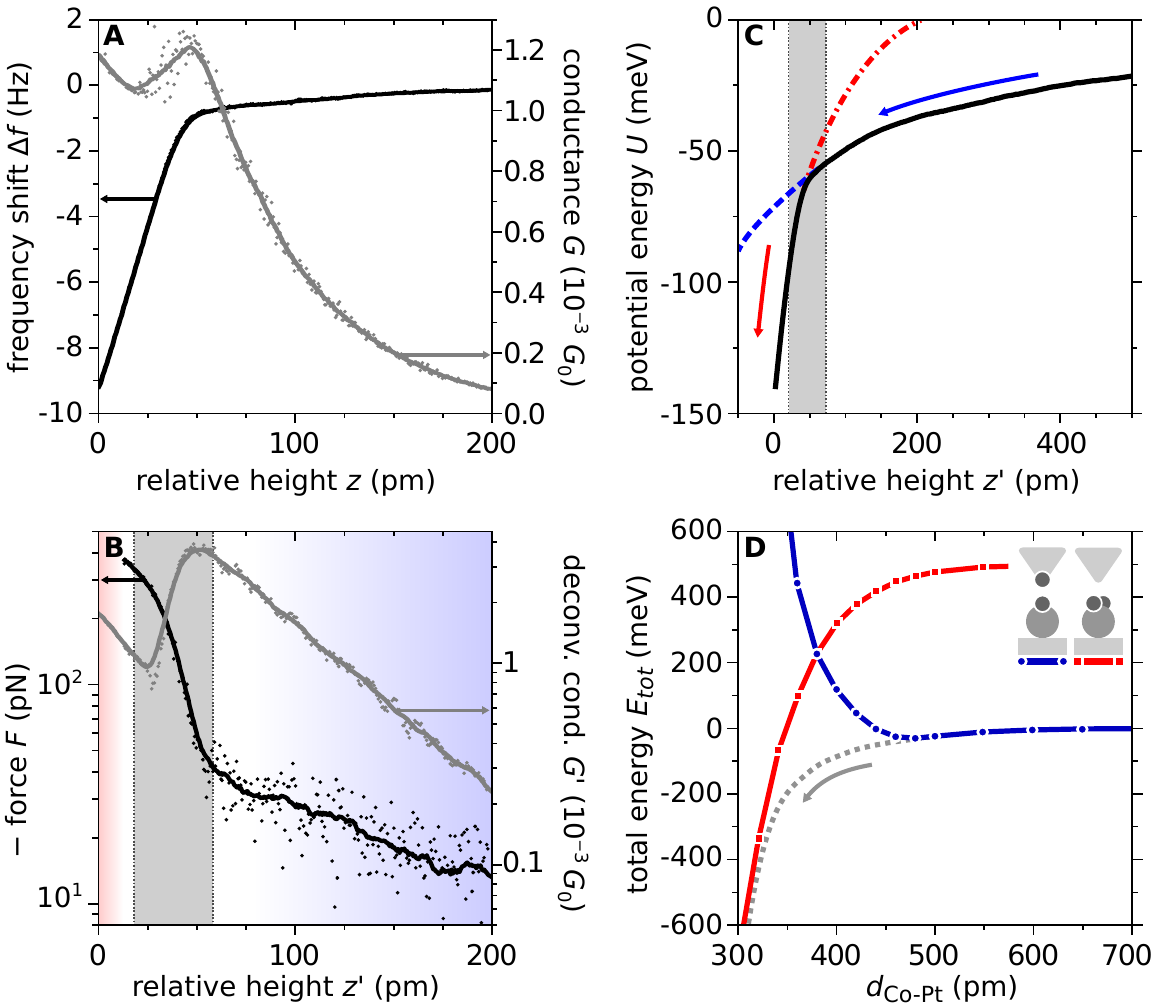}}
  		\caption{\textbf{Force Measurements on a Switching Complex.} (A) Simultaneous frequency shift-distance, \var{\Delta f(z)} (black), and conductance-distance, \var{G(z)} (gray), measurements on a \chem{CoH} $\var{S} = 1$ complex with hydrogen-functionalized tip. The spin transition, occurring at a relative \var{z} height of $50~\unit{pm}$, is evident in both force and conductance channels. (B) Frequency shift was converted to short-range forces (black) and the conductance was deconvoluted to remove averaging over the oscillation amplitude (gray). On either side of the transition region the deconvoluted conductance and force increase exponentially and can be described by the expressions, $G(z) = G_0 \exp (-2\kappa_G\, (z_0 + z))$ and $F(z') = F_0 \exp (-2 \kappa_F \, (z_0+z'))$, respectively. Inverse decay constants, $\var{\kappa_G}\,[\unit{nm^{-1}}]$: $13.0 \pm 0.5$ $(0 < \var{z'} < 30~\unit{pm})$, $9.5 \pm 0.1$ $(70 < \var{z'} < 200~\unit{pm})$; $\kappa_F\,[\unit{nm^{-1}}]$: $10.0 \pm 0.5$ $(0 < \var{z'} < 30~\unit{pm})$, $4.2 \pm 0.3$ $(70 < \var{z'} < 200~\unit{pm})$. (C) Interaction potential energy surface during the $\var{S} = 1$ to $\var{S} = \half$ transition (black) determined by integrating the experimental \var{F(z')} data. Dashed lines highlight the change in slope and indicate the point where a lower potential energy surface becomes accessible. Vertical dashed lines in (B) and (C) indicate the transition regime. For all curves, zero distance corresponds to the point of closest approach. (D) Simulated diabatic potential energy curves for a \chem{CoH/\hBN/Rh(111)} complex approached by a \chem{H}-functionalized \chem{Pt} tip (dashed blue) and a \chem{CoH_2} approached with a bare tip (dash-dotted red). The approximate adiabatic curve is shown as grey dotted line. The reaction coordinate \var{d_{\chem{Co-Pt}}} is the distance between the \chem{Co} and the apex \chem{Pt} atoms.}
		\label{fig:03}
	\end{figure*}

	To reveal the microscopic forces at work in the spin transition, we track the frequency shift, $\var{\Delta f}$, of the oscillating tuning fork from its non-interacting resonance frequency $f_0 = 29,077~\unit{Hz}$. We measure \var{\Delta f(z)} curves over switching complexes and the bare \chem{\hBN}. To remove the long-range forces between the extended tip and the sample we subtract the background from the data, \textit{i.e.} $\var{\Delta f} = \var{\Delta f_{\chem{CoH}}} - \var{\Delta f_{\chem{\hBN}}}$ (see Methods and Fig.~\ref{fig:S02}). The \var{\Delta f} is small and negative before rapidly decreasing on approach (Figure~\ref{fig:03}A, black). This sharp drop in \var{\Delta f} coincides with a change in the \var{G(z)} measurement similar to Figure~\ref{fig:02}B, however here this feature is broadened due to averaging over the $100~\unit{pm}$ oscillation amplitude (Figure~\ref{fig:03}A, gray). Short range forces, \var{F(z')}, were quantified by converting \var{\Delta f(z)} using the method of Sader and Jarvis.~\cite{Sader2004} Prior to the $\var{S} = 1$ to $\var{S} = \half$ transition, the force between the tip and sample is weakly attractive and grows exponentially upon approach. As the hydrogen on the tip apex couples to the \chem{CoH}-complex, the attractive force grows steeply over a transition region of $35~\unit{pm}$ before leveling off  (Figure~\ref{fig:03}B). The instantaneous junction conductance, \var{G(z')}, deconvoluted to remove the influence of an oscillating tip,~\cite{Sader2010} reveals that the force and conductance transition regions coincide.

	From the \var{F(z')} measurement, we reconstruct the 1D potential energy landscape, \var{U(z')}, across the spin transition by integrating \var{F(z')} (Figure~\ref{fig:03}C). The \var{U(z')} curve shows a steep change in slope as the tip brings the hydrogen closer to the \chem{CoH} complex, suggestive of a transition between potential energy surfaces. This interpretation is broadly in line with a framework recently developed by Hapala \textit{et al.} to describe high-resolution AFM imaging with functionalized tips.~\cite{Hapala2014} One key component of this model is that the probe particle, hydrogen in our case, not only follows the lowest potential energy surface but also undergoes relaxation within the junction. In our \var{U(z')} measurement, the kink corresponds to the relative $z'$ distance where the hydrogen on the tip apex can relax onto a lower potential energy surface. Chemically, the \chem{CoH} is transformed into \chem{CoH_2}, a magnetically distinct complex with a different potential energy surface. DFT calculations show that the transition between the two distinct chemical configurations proceeds via a continuum of intermediate transition states. This intermediate regime is characterized by hydrogen on the tip sharing its bond with the tip apex and the \chem{CoH} while simultaneously undergoing reorientation within the junction. To tackle this complexity, it is instructive to examine the diabatic potential energy surfaces in the limiting cases, that is, when \chem{CoH} interacts with a hydrogen-functionalized tip and \chem{CoH_2} interacting with a clean metallic tip. In Figure~\ref{fig:03}D we plot the dependence of the total energies of the two above configurations (dash-dotted red and dashed blue curves, respectively) on the tip-sample distance $d$ (see methods section and Figure~\ref{fig:S06}), the chosen reaction coordinate for our system. The intersection of the two potential energy curves with varying junction width confirms that hydrogen transfer drives the spin transition and is consistent with experimental observations. As mentioned above, the system does not follow the diabatic curves, rather, it undergoes a barrierless transition along a reaction path difficult to precisely identify by \textit{ab initio} methods due to the shallowness and unevenness of the hydrogen potential energy landscape in the junction. An approximate shape of this pathway is shown in Figure~\ref{fig:03}D with a grey dashed line linking the two asymptotic cases. The presence of a transition region (grey regions in Figure~\ref{fig:03}B,C) further indicates that the spin switching is driven by the approaching hydrogen and that the residence time is shorter than the measurement time, leading to an averaging and smearing-out when probed with the oscillating tip. Furthermore, the increased mechanical dissipation during the transition ($\sim55~\unit{meV/cycle}$, see Figure~\ref{fig:S03}) is in good agreement with the potential energy difference across the transition region of the \var{U(z')} measurement ($\sim35~\unit{meV}$) and points to the existence of hysteresis on the timescale of the oscillation period $(2\var{f_0})^{-1} \sim 15~\unit{\mu s}$.
	
\section*{Discussion}

	We have reversibly switched the spin state of cobalt hydride complexes in a tunnel junction by using the hydrogen on the tip apex as a tunable ligand.  During the total spin change, a transition region exists where the hydrogen can occupy two nearly equivalent sites separated only by a negligible barrier.  As the junction conductance is altered, site equivalence is removed and the tip displacement shifts the system into a preferred site.  Beyond total spin, interactions with the tip determine how the cobalt spin couples with the host electron bath and whether a correlated Kondo state emerges.  The relative stability of the double potential well and the possible occurrence of correlations is expected to be dependent on the materials used and the reactivity of the adsorbed molecule on the tip and the undercoordinated adatom.  Reconstructing the 1D potential energy landscape shows that chemical transformation within the junction is responsible for the change of the total spin.  Our measurements highlight how tip functionalization can influence the spin under investigation and suggests that they may be used to tailor molecular spins in ways difficult to achieve through traditional chemical synthesis.
	
\section*{Materials and Methods}

	The clean \chem{Rh(111)} surface was prepared by multiple cycles of argon ion sputtering and annealing to $1100~\unit{K}$ in an ultrahigh vacuum chamber. During the final annealing cycle borazine (\chem{B_3N_3H_6}) gas was exposed to the hot sample at a pressure of $1.2\times10^{-6}~\unit{mbar}$ for about $\sim2$ minutes, resulting in a monoatomic layer of \hBN. Subsequently, the sample was cooled down ($\sim20\unit{K}$) and cobalt was deposited via an electron beam evaporator. As hydrogen is the predominant component of the residual gas background it is responsible for the formation of the cobalt hydride complexes. STM/AFM experiments were performed on a homebuilt instrument operating in ultra-high vacuum at a base temperature of $1.1~\unit{K}$. All spectroscopic (\var{dI/dV}) measurements were obtained by adding a small sinusoidal voltage ($\var{V_{mod}} = 0.2~\unit{mV}$, $\var{f_{mod}} = 600-800~\unit{Hz}$) to the bias voltage \var{V} and using an external lock-in amplifier. We used $\var{V} = -10~\unit{mV}$ for adjusting the conductance \var{G} prior recording \var{dI/dV} curves by varying \var{V} at constant \var{z}, as well as for recording \var{G(z)} and \var{\Delta f(z)} curves at constant \var{V}. Hydrogen pickup occurs fortuitously while performing our experiments, for example, during the approach of an initially \chem{CoH_x} (\chem{x}=1--2) complex with a non-functionalized tip. Hydrogen-terminated tips were initially identified by features such as enhanced contrast in constant current imaging and then verified by the characteristic non-exponential behavior in \var{G(z)} measurements (Figure~\ref{fig:02}B, green).

	The quartz tuning fork has a resonance frequency of $\var{f_0} = 29,077~\unit{Hz}$ and a quality factor of approximately 10,000. Oscillation amplitudes of $100~\unit{pm}$ were used. For the frequency shift curves, \var{\Delta f(z)}, the tip was positioned above the \chem{CoH} complex with the oscillating tuning fork at $\var{G} = 1.72 \times 10^{-5}~\var{G_0}$. The feedback loop was disabled and the oscillating tip then approached $200-300~\unit{pm}$ towards the sample and back. Immediately after completion, a retract curve was obtained by moving $2000~\unit{pm}$ away from the surface (see Figure~\ref{fig:S02}). When this sequence was completed, the tip was moved at constant height to the bare \chem{\hBN} and \var{\Delta f_{\chem{\hBN}}} was obtained. The frequency shift due to short-range forces are obtained by taking the difference $\var{\Delta f} = \var{\Delta f_{\chem{CoH}}} - \var{\Delta f_{\chem{\hBN}}}$. We distinguish between time averaged signals at tip heights $z$ and deconvoluted (instantaneous) signals at tip heights \var{z'}. 

	First principles calculations have been carried out in a manner consistent with the calculations employed in our previous work. ~\cite{Jacobson2015} In brief, a density functional theory approach, based on the projector-augmented-wave method~\cite{Blochl1994} and a plane-wave basis set,~\cite{Kresse1996} was used as implemented in the Vienna Ab-initio Simulation Package (VASP).~\cite{Kresse1993,Kresse1996} Exchange and correlation were treated with the gradient-corrected functional as formalized by Perdew, Burke and Ernzerhof.~\cite{Perdew1996} On-site Coulomb interaction corrections were accounted for in the framework of the LSDA+U formalism as introduced by Dudarev \textit{et al.}~\cite{Dudarev1998} The considered geometry of the system was identical to the one introduced in Ref.~\onlinecite{Jacobson2015}, \textit{i.e.} a \chem{Rh(111)} surface was simulated by 5 \chem{Rh} layers on top of which an \chem{\hBN} sheet was deposited (considered to be commensurate for computational feasibility reasons). A \chem{CoH} complex was residing on top of an N atom of the \chem{\hBN} sheet. The tip was simulated by a \chem{Pt} pyramid of 4 atoms attached to a \chem{Rh} surface. Prior to the tip approach simulation, the tip and the sample were allowed to relax fully. After placing the tip in the vicinity of the \chem{CoH_x} complex, the \chem{CoH_x} complex and the apex atom of the tip (and the attached hydrogen atom, where appropriate) were allowed to relax again, assuming an equilibrium static configuration. A full relaxation of the tip and the \hBN atoms around the \chem{CoH_x} complex was tested and not found to introduce any change to the results above the calculation error level. In the experimentally relevant case of a \chem{CoH} complex being approached with a singly hydrogenated tip we are thus left with four atoms allowed to relax in 3 dimensions amounting to a 12-dimensional configuration space. In effect, however, both the tip apex Pt and the Co atoms were found to experience only vertical relaxations which, according to our tests, result in minor quantitative change of the total energies and magnetic moments. This complies fully with our understanding of the mechanism of \chem{CoH_x} complex spin switching, the latter being governed fully by the \chem{Co-H} bonding, which in turn is altered through the rearrangement of the hydrogen atoms. To underline this fact, the calculation results presented in Figures~\ref{fig:03}D and \ref{fig:S04} were obtained in a diabatic fashion, meaning that the initial relative orientation of the \chem{CoH_x} and the \chem{TipH_{2-x}} atoms were kept frozen and only the distance between the tip and the sample (characterized by the distance $d$ between \chem{Co} and apex \chem{Pt} atoms - see Figure~\ref{fig:S03}) was varied. For hydrogen adsorption and transfer energies as well as the adiabatic transition calculation the junction was allowed to relax fully.

\section*{Acknowledgements}
	The authors thank U. Schlickum for fruitful discussions, F. Giessibl for providing quartz tuning forks, and M. Mausser for focused ion beam cutting of the Pt tips.  
	
	\textbf{Funding:} P.J. acknowledges support from the Alexander von Humboldt Foundation.  M.M. and M.T. acknowledge support from the SFB 767.  O.B. and V.S. acknowledge support from the SFB 762. O.B. acknowledges support from the ERC Grant No. 320796, MODPHYSFRICT.  
	
	\textbf{Author contributions:}  M.T. and K.K. conceived the project.  P.J., M.M., and G.L. performed the measurements.  P.J., M.M., and M.T. analyzed the data.  O.B. and V.S. performed the density functional theory calculations. P.J. drafted the manuscript, all authors contributed to the manuscript. 
	
	\textbf{Competing interests:} None
	

\begin{thebibliography}{31}
\expandafter\ifx\csname natexlab\endcsname\relax\def\natexlab#1{#1}\fi
\expandafter\ifx\csname bibnamefont\endcsname\relax
  \def\bibnamefont#1{#1}\fi
\expandafter\ifx\csname bibfnamefont\endcsname\relax
  \def\bibfnamefont#1{#1}\fi
\expandafter\ifx\csname citenamefont\endcsname\relax
  \def\citenamefont#1{#1}\fi
\expandafter\ifx\csname url\endcsname\relax
  \def\url#1{\texttt{#1}}\fi
\expandafter\ifx\csname urlprefix\endcsname\relax\def\urlprefix{URL }\fi
\providecommand{\bibinfo}[2]{#2}
\providecommand{\eprint}[2][]{\url{#2}}

\bibitem[{\citenamefont{Otte et~al.}(2008)\citenamefont{Otte, Ternes, von
  Bergmann, Loth, Brune, Lutz, Hirjibehedin, and Heinrich}}]{Otte2008}
\bibinfo{author}{\bibfnamefont{A.~F.} \bibnamefont{Otte}},
  \bibinfo{author}{\bibfnamefont{M.}~\bibnamefont{Ternes}},
  \bibinfo{author}{\bibfnamefont{K.}~\bibnamefont{von Bergmann}},
  \bibinfo{author}{\bibfnamefont{S.}~\bibnamefont{Loth}},
  \bibinfo{author}{\bibfnamefont{H.}~\bibnamefont{Brune}},
  \bibinfo{author}{\bibfnamefont{C.~P.} \bibnamefont{Lutz}},
  \bibinfo{author}{\bibfnamefont{C.~F.} \bibnamefont{Hirjibehedin}},
  \bibnamefont{and} \bibinfo{author}{\bibfnamefont{A.~J.}
  \bibnamefont{Heinrich}}, \bibinfo{journal}{Nature Physics}
  \textbf{\bibinfo{volume}{4}}, \bibinfo{pages}{847} (\bibinfo{year}{2008}),
  ISSN \bibinfo{issn}{1745-2473},
  \urlprefix\url{http://www.nature.com/doifinder/10.1038/nphys1072}.

\bibitem[{\citenamefont{Kahle et~al.}(2012)\citenamefont{Kahle, Deng,
  Malinowski, Tonnoir, Forment-Aliaga, Thontasen, Rinke, Le, Turkowski, Rahman
  et~al.}}]{Kahle2011}
\bibinfo{author}{\bibfnamefont{S.}~\bibnamefont{Kahle}},
  \bibinfo{author}{\bibfnamefont{Z.}~\bibnamefont{Deng}},
  \bibinfo{author}{\bibfnamefont{N.}~\bibnamefont{Malinowski}},
  \bibinfo{author}{\bibfnamefont{C.}~\bibnamefont{Tonnoir}},
  \bibinfo{author}{\bibfnamefont{A.}~\bibnamefont{Forment-Aliaga}},
  \bibinfo{author}{\bibfnamefont{N.}~\bibnamefont{Thontasen}},
  \bibinfo{author}{\bibfnamefont{G.}~\bibnamefont{Rinke}},
  \bibinfo{author}{\bibfnamefont{D.}~\bibnamefont{Le}},
  \bibinfo{author}{\bibfnamefont{V.}~\bibnamefont{Turkowski}},
  \bibinfo{author}{\bibfnamefont{T.~S.} \bibnamefont{Rahman}},
  \bibnamefont{et~al.}, \bibinfo{journal}{Nano Letters}
  \textbf{\bibinfo{volume}{12}}, \bibinfo{pages}{518} (\bibinfo{year}{2012}),
  ISSN \bibinfo{issn}{1530-6984},
  \urlprefix\url{http://pubs.acs.org/doi/abs/10.1021/nl204141z}.

\bibitem[{\citenamefont{Rau et~al.}(2014)\citenamefont{Rau, Baumann, Rusponi,
  Donati, Stepanow, Gragnaniello, Dreiser, Piamonteze, Nolting, Gangopadhyay
  et~al.}}]{Rau2014}
\bibinfo{author}{\bibfnamefont{I.~G.} \bibnamefont{Rau}},
  \bibinfo{author}{\bibfnamefont{S.}~\bibnamefont{Baumann}},
  \bibinfo{author}{\bibfnamefont{S.}~\bibnamefont{Rusponi}},
  \bibinfo{author}{\bibfnamefont{F.}~\bibnamefont{Donati}},
  \bibinfo{author}{\bibfnamefont{S.}~\bibnamefont{Stepanow}},
  \bibinfo{author}{\bibfnamefont{L.}~\bibnamefont{Gragnaniello}},
  \bibinfo{author}{\bibfnamefont{J.}~\bibnamefont{Dreiser}},
  \bibinfo{author}{\bibfnamefont{C.}~\bibnamefont{Piamonteze}},
  \bibinfo{author}{\bibfnamefont{F.}~\bibnamefont{Nolting}},
  \bibinfo{author}{\bibfnamefont{S.}~\bibnamefont{Gangopadhyay}},
  \bibnamefont{et~al.}, \bibinfo{journal}{Science}
  \textbf{\bibinfo{volume}{344}}, \bibinfo{pages}{988} (\bibinfo{year}{2014}),
  ISSN \bibinfo{issn}{0036-8075},
  \urlprefix\url{http://www.sciencemag.org/cgi/doi/10.1126/science.1252841
  http://www.ncbi.nlm.nih.gov/pubmed/24812206}.

\bibitem[{\citenamefont{Oberg et~al.}(2014)\citenamefont{Oberg, Calvo, Delgado,
  Moro-Lagares, Serrate, Jacob, Fern{\'{a}}ndez-Rossier, and
  Hirjibehedin}}]{Oberg2014}
\bibinfo{author}{\bibfnamefont{J.~C.} \bibnamefont{Oberg}},
  \bibinfo{author}{\bibfnamefont{M.~R.} \bibnamefont{Calvo}},
  \bibinfo{author}{\bibfnamefont{F.}~\bibnamefont{Delgado}},
  \bibinfo{author}{\bibfnamefont{M.}~\bibnamefont{Moro-Lagares}},
  \bibinfo{author}{\bibfnamefont{D.}~\bibnamefont{Serrate}},
  \bibinfo{author}{\bibfnamefont{D.}~\bibnamefont{Jacob}},
  \bibinfo{author}{\bibfnamefont{J.}~\bibnamefont{Fern{\'{a}}ndez-Rossier}},
  \bibnamefont{and} \bibinfo{author}{\bibfnamefont{C.~F.}
  \bibnamefont{Hirjibehedin}}, \bibinfo{journal}{Nature Nanotechnology}
  \textbf{\bibinfo{volume}{9}}, \bibinfo{pages}{64} (\bibinfo{year}{2014}),
  ISSN \bibinfo{issn}{1748-3387},
  \urlprefix\url{http://www.ncbi.nlm.nih.gov/pubmed/24317285
  http://www.nature.com/doifinder/10.1038/nnano.2013.264}.

\bibitem[{\citenamefont{Jacobson et~al.}(2015)\citenamefont{Jacobson, Herden,
  Muenks, Laskin, Brovko, Stepanyuk, Ternes, and Kern}}]{Jacobson2015}
\bibinfo{author}{\bibfnamefont{P.}~\bibnamefont{Jacobson}},
  \bibinfo{author}{\bibfnamefont{T.}~\bibnamefont{Herden}},
  \bibinfo{author}{\bibfnamefont{M.}~\bibnamefont{Muenks}},
  \bibinfo{author}{\bibfnamefont{G.}~\bibnamefont{Laskin}},
  \bibinfo{author}{\bibfnamefont{O.}~\bibnamefont{Brovko}},
  \bibinfo{author}{\bibfnamefont{V.}~\bibnamefont{Stepanyuk}},
  \bibinfo{author}{\bibfnamefont{M.}~\bibnamefont{Ternes}}, \bibnamefont{and}
  \bibinfo{author}{\bibfnamefont{K.}~\bibnamefont{Kern}},
  \bibinfo{journal}{Nature Communications} \textbf{\bibinfo{volume}{6}},
  \bibinfo{pages}{8536} (\bibinfo{year}{2015}), ISSN \bibinfo{issn}{2041-1723},
  \urlprefix\url{http://arxiv.org/abs/1505.02277
  http://www.nature.com/doifinder/10.1038/ncomms9536}.

\bibitem[{\citenamefont{Heinrich et~al.}(2015)\citenamefont{Heinrich, Braun,
  Pascual, and Franke}}]{Heinrich2015}
\bibinfo{author}{\bibfnamefont{B.~W.} \bibnamefont{Heinrich}},
  \bibinfo{author}{\bibfnamefont{L.}~\bibnamefont{Braun}},
  \bibinfo{author}{\bibfnamefont{J.~I.} \bibnamefont{Pascual}},
  \bibnamefont{and} \bibinfo{author}{\bibfnamefont{K.~J.}
  \bibnamefont{Franke}}, \bibinfo{journal}{Nano Letters}
  \textbf{\bibinfo{volume}{15}}, \bibinfo{pages}{4024} (\bibinfo{year}{2015}),
  ISSN \bibinfo{issn}{1530-6984},
  \urlprefix\url{http://pubs.acs.org/doi/abs/10.1021/acs.nanolett.5b00987}.

\bibitem[{\citenamefont{Parks et~al.}(2010)\citenamefont{Parks, Champagne,
  Costi, Shum, Pasupathy, Neuscamman, Flores-Torres, Cornaglia, Aligia,
  Balseiro et~al.}}]{Parks2010}
\bibinfo{author}{\bibfnamefont{J.~J.} \bibnamefont{Parks}},
  \bibinfo{author}{\bibfnamefont{A.~R.} \bibnamefont{Champagne}},
  \bibinfo{author}{\bibfnamefont{T.~A.} \bibnamefont{Costi}},
  \bibinfo{author}{\bibfnamefont{W.~W.} \bibnamefont{Shum}},
  \bibinfo{author}{\bibfnamefont{A.~N.} \bibnamefont{Pasupathy}},
  \bibinfo{author}{\bibfnamefont{E.}~\bibnamefont{Neuscamman}},
  \bibinfo{author}{\bibfnamefont{S.}~\bibnamefont{Flores-Torres}},
  \bibinfo{author}{\bibfnamefont{P.~S.} \bibnamefont{Cornaglia}},
  \bibinfo{author}{\bibfnamefont{A.~A.} \bibnamefont{Aligia}},
  \bibinfo{author}{\bibfnamefont{C.~A.} \bibnamefont{Balseiro}},
  \bibnamefont{et~al.}, \bibinfo{journal}{Science}
  \textbf{\bibinfo{volume}{328}}, \bibinfo{pages}{1370} (\bibinfo{year}{2010}),
  ISSN \bibinfo{issn}{0036-8075},
  \urlprefix\url{http://www.sciencemag.org/cgi/doi/10.1126/science.1186874}.

\bibitem[{\citenamefont{Eigler and Schweizer}(1990)}]{Eigler1990}
\bibinfo{author}{\bibfnamefont{D.~M.} \bibnamefont{Eigler}} \bibnamefont{and}
  \bibinfo{author}{\bibfnamefont{E.~K.} \bibnamefont{Schweizer}},
  \bibinfo{journal}{Nature} \textbf{\bibinfo{volume}{344}},
  \bibinfo{pages}{524} (\bibinfo{year}{1990}), ISSN \bibinfo{issn}{0028-0836},
  \urlprefix\url{http://www.nature.com/doifinder/10.1038/344524a0}.

\bibitem[{\citenamefont{Eigler et~al.}(1991)\citenamefont{Eigler, Lutz, and
  Rudge}}]{Eigler1991}
\bibinfo{author}{\bibfnamefont{D.~M.} \bibnamefont{Eigler}},
  \bibinfo{author}{\bibfnamefont{C.~P.} \bibnamefont{Lutz}}, \bibnamefont{and}
  \bibinfo{author}{\bibfnamefont{W.~E.} \bibnamefont{Rudge}},
  \bibinfo{journal}{Nature} \textbf{\bibinfo{volume}{352}},
  \bibinfo{pages}{600} (\bibinfo{year}{1991}), ISSN \bibinfo{issn}{0028-0836},
  \urlprefix\url{http://www.nature.com/doifinder/10.1038/352600a0}.

\bibitem[{\citenamefont{Custance et~al.}(2009)\citenamefont{Custance, Perez,
  and Morita}}]{Custance2009}
\bibinfo{author}{\bibfnamefont{O.}~\bibnamefont{Custance}},
  \bibinfo{author}{\bibfnamefont{R.}~\bibnamefont{Perez}}, \bibnamefont{and}
  \bibinfo{author}{\bibfnamefont{S.}~\bibnamefont{Morita}},
  \bibinfo{journal}{Nature nanotechnology} \textbf{\bibinfo{volume}{4}},
  \bibinfo{pages}{803} (\bibinfo{year}{2009}), ISSN \bibinfo{issn}{1748-3387},
  \urlprefix\url{http://dx.doi.org/10.1038/nnano.2009.347}.

\bibitem[{\citenamefont{Bork et~al.}(2011)\citenamefont{Bork, Zhang,
  Diekh{\"{o}}ner, Borda, Simon, Kroha, Wahl, and Kern}}]{Bork2011}
\bibinfo{author}{\bibfnamefont{J.}~\bibnamefont{Bork}},
  \bibinfo{author}{\bibfnamefont{Y.-H.} \bibnamefont{Zhang}},
  \bibinfo{author}{\bibfnamefont{L.}~\bibnamefont{Diekh{\"{o}}ner}},
  \bibinfo{author}{\bibfnamefont{L.}~\bibnamefont{Borda}},
  \bibinfo{author}{\bibfnamefont{P.}~\bibnamefont{Simon}},
  \bibinfo{author}{\bibfnamefont{J.}~\bibnamefont{Kroha}},
  \bibinfo{author}{\bibfnamefont{P.}~\bibnamefont{Wahl}}, \bibnamefont{and}
  \bibinfo{author}{\bibfnamefont{K.}~\bibnamefont{Kern}},
  \bibinfo{journal}{Nature Physics} \textbf{\bibinfo{volume}{7}},
  \bibinfo{pages}{22} (\bibinfo{year}{2011}), ISSN \bibinfo{issn}{1745-2473},
  \urlprefix\url{http://arxiv.org/abs/1108.0869}.

\bibitem[{\citenamefont{Choi et~al.}(2012)\citenamefont{Choi, Rastei, Simon,
  and Limot}}]{Choi2012}
\bibinfo{author}{\bibfnamefont{D.~J.} \bibnamefont{Choi}},
  \bibinfo{author}{\bibfnamefont{M.~V.} \bibnamefont{Rastei}},
  \bibinfo{author}{\bibfnamefont{P.}~\bibnamefont{Simon}}, \bibnamefont{and}
  \bibinfo{author}{\bibfnamefont{L.}~\bibnamefont{Limot}},
  \bibinfo{journal}{Physical Review Letters} \textbf{\bibinfo{volume}{108}},
  \bibinfo{pages}{1} (\bibinfo{year}{2012}), ISSN \bibinfo{issn}{00319007}.

\bibitem[{\citenamefont{Muenks et~al.}(2016)\citenamefont{Muenks, Jacobson,
  Ternes, and Kern}}]{Muenks2016}
\bibinfo{author}{\bibfnamefont{M.}~\bibnamefont{Muenks}},
  \bibinfo{author}{\bibfnamefont{P.}~\bibnamefont{Jacobson}},
  \bibinfo{author}{\bibfnamefont{M.}~\bibnamefont{Ternes}}, \bibnamefont{and}
  \bibinfo{author}{\bibfnamefont{K.}~\bibnamefont{Kern}}
  (\bibinfo{year}{2016}), \eprint{1605.02798v1}.

\bibitem[{\citenamefont{Wagner and Temirov}(2015)}]{Wagner2015}
\bibinfo{author}{\bibfnamefont{C.}~\bibnamefont{Wagner}} \bibnamefont{and}
  \bibinfo{author}{\bibfnamefont{R.}~\bibnamefont{Temirov}},
  \bibinfo{journal}{Progress in Surface Science} \textbf{\bibinfo{volume}{90}},
  \bibinfo{pages}{194} (\bibinfo{year}{2015}), ISSN \bibinfo{issn}{00796816},
  \urlprefix\url{http://dx.doi.org/10.1016/j.progsurf.2015.01.001
  http://linkinghub.elsevier.com/retrieve/pii/S0079681615000027}.

\bibitem[{\citenamefont{Bartels et~al.}(1997)\citenamefont{Bartels, Meyer, and
  Rieder}}]{Bartels1997}
\bibinfo{author}{\bibfnamefont{L.}~\bibnamefont{Bartels}},
  \bibinfo{author}{\bibfnamefont{G.}~\bibnamefont{Meyer}}, \bibnamefont{and}
  \bibinfo{author}{\bibfnamefont{K.-H.} \bibnamefont{Rieder}},
  \bibinfo{journal}{Applied Physics Letters} \textbf{\bibinfo{volume}{71}},
  \bibinfo{pages}{213} (\bibinfo{year}{1997}), ISSN \bibinfo{issn}{00036951},
  
\urlprefix\url{
http://scitation.aip.org/content/aip/journal/apl/71/2/10.1063/1.119503}.

\bibitem[{\citenamefont{Weymouth et~al.}(2014)\citenamefont{Weymouth, Hofmann,
  and Giessibl}}]{Weymouth2014}
\bibinfo{author}{\bibfnamefont{A.~J.} \bibnamefont{Weymouth}},
  \bibinfo{author}{\bibfnamefont{T.}~\bibnamefont{Hofmann}}, \bibnamefont{and}
  \bibinfo{author}{\bibfnamefont{F.~J.} \bibnamefont{Giessibl}},
  \bibinfo{journal}{Science} \textbf{\bibinfo{volume}{343}},
  \bibinfo{pages}{1120} (\bibinfo{year}{2014}), ISSN \bibinfo{issn}{1095-9203},
  \urlprefix\url{http://www.sciencemag.org/content/343/6175/1120}.

\bibitem[{\citenamefont{Kichin et~al.}(2011)\citenamefont{Kichin, Weiss,
  Wagner, Tautz, and Temirov}}]{Kichin2011}
\bibinfo{author}{\bibfnamefont{G.}~\bibnamefont{Kichin}},
  \bibinfo{author}{\bibfnamefont{C.}~\bibnamefont{Weiss}},
  \bibinfo{author}{\bibfnamefont{C.}~\bibnamefont{Wagner}},
  \bibinfo{author}{\bibfnamefont{F.~S.} \bibnamefont{Tautz}}, \bibnamefont{and}
  \bibinfo{author}{\bibfnamefont{R.}~\bibnamefont{Temirov}},
  \bibinfo{journal}{Journal of the American Chemical Society}
  \textbf{\bibinfo{volume}{133}}, \bibinfo{pages}{16847}
  (\bibinfo{year}{2011}), ISSN \bibinfo{issn}{0002-7863},
  \urlprefix\url{http://pubs.acs.org/doi/abs/10.1021/ja204624g}.

\bibitem[{\citenamefont{Weiss et~al.}(2010)\citenamefont{Weiss, Wagner,
  Kleimann, Rohlfing, Tautz, and Temirov}}]{Weiss2010}
\bibinfo{author}{\bibfnamefont{C.}~\bibnamefont{Weiss}},
  \bibinfo{author}{\bibfnamefont{C.}~\bibnamefont{Wagner}},
  \bibinfo{author}{\bibfnamefont{C.}~\bibnamefont{Kleimann}},
  \bibinfo{author}{\bibfnamefont{M.}~\bibnamefont{Rohlfing}},
  \bibinfo{author}{\bibfnamefont{F.~S.} \bibnamefont{Tautz}}, \bibnamefont{and}
  \bibinfo{author}{\bibfnamefont{R.}~\bibnamefont{Temirov}},
  \bibinfo{journal}{Physical Review Letters} \textbf{\bibinfo{volume}{105}},
  \bibinfo{pages}{2} (\bibinfo{year}{2010}), ISSN \bibinfo{issn}{00319007}.

\bibitem[{\citenamefont{Wende et~al.}(2007)\citenamefont{Wende, Bernien, Luo,
  Sorg, Ponpandian, Kurde, Miguel, Piantek, Xu, Eckhold et~al.}}]{Wende2007}
\bibinfo{author}{\bibfnamefont{H.}~\bibnamefont{Wende}},
  \bibinfo{author}{\bibfnamefont{M.}~\bibnamefont{Bernien}},
  \bibinfo{author}{\bibfnamefont{J.}~\bibnamefont{Luo}},
  \bibinfo{author}{\bibfnamefont{C.}~\bibnamefont{Sorg}},
  \bibinfo{author}{\bibfnamefont{N.}~\bibnamefont{Ponpandian}},
  \bibinfo{author}{\bibfnamefont{J.}~\bibnamefont{Kurde}},
  \bibinfo{author}{\bibfnamefont{J.}~\bibnamefont{Miguel}},
  \bibinfo{author}{\bibfnamefont{M.}~\bibnamefont{Piantek}},
  \bibinfo{author}{\bibfnamefont{X.}~\bibnamefont{Xu}},
  \bibinfo{author}{\bibfnamefont{P.}~\bibnamefont{Eckhold}},
  \bibnamefont{et~al.}, \bibinfo{journal}{Nature Materials}
  \textbf{\bibinfo{volume}{6}}, \bibinfo{pages}{516} (\bibinfo{year}{2007}),
  ISSN \bibinfo{issn}{1476-1122},
  
\urlprefix\url{
http://www.ncbi.nlm.nih.gov/pubmed/17558431$\backslash$nhttp://www.nature.com/nm
at/journal/v6/n7/full/nmat1932.html
  http://www.nature.com/doifinder/10.1038/nmat1932}.

\bibitem[{\citenamefont{W{\"{a}}ckerlin
  et~al.}(2012)\citenamefont{W{\"{a}}ckerlin, Tarafder, Siewert, Girovsky,
  H{\"{a}}hlen, Iacovita, Kleibert, Nolting, Jung, Oppeneer
  et~al.}}]{Wackerlin2012}
\bibinfo{author}{\bibfnamefont{C.}~\bibnamefont{W{\"{a}}ckerlin}},
  \bibinfo{author}{\bibfnamefont{K.}~\bibnamefont{Tarafder}},
  \bibinfo{author}{\bibfnamefont{D.}~\bibnamefont{Siewert}},
  \bibinfo{author}{\bibfnamefont{J.}~\bibnamefont{Girovsky}},
  \bibinfo{author}{\bibfnamefont{T.}~\bibnamefont{H{\"{a}}hlen}},
  \bibinfo{author}{\bibfnamefont{C.}~\bibnamefont{Iacovita}},
  \bibinfo{author}{\bibfnamefont{A.}~\bibnamefont{Kleibert}},
  \bibinfo{author}{\bibfnamefont{F.}~\bibnamefont{Nolting}},
  \bibinfo{author}{\bibfnamefont{T.~A.} \bibnamefont{Jung}},
  \bibinfo{author}{\bibfnamefont{P.~M.} \bibnamefont{Oppeneer}},
  \bibnamefont{et~al.}, \bibinfo{journal}{Chemical Science}
  \textbf{\bibinfo{volume}{3}}, \bibinfo{pages}{3154} (\bibinfo{year}{2012}),
  ISSN \bibinfo{issn}{2041-6520},
  \urlprefix\url{http://xlink.rsc.org/?DOI=c1jm11679g
  http://xlink.rsc.org/?DOI=c2sc20828h}.

\bibitem[{\citenamefont{Temirov et~al.}(2008)\citenamefont{Temirov, Soubatch,
  Neucheva, Lassise, and Tautz}}]{Temirov2008}
\bibinfo{author}{\bibfnamefont{R.}~\bibnamefont{Temirov}},
  \bibinfo{author}{\bibfnamefont{S.}~\bibnamefont{Soubatch}},
  \bibinfo{author}{\bibfnamefont{O.}~\bibnamefont{Neucheva}},
  \bibinfo{author}{\bibfnamefont{A.~C.} \bibnamefont{Lassise}},
  \bibnamefont{and} \bibinfo{author}{\bibfnamefont{F.~S.} \bibnamefont{Tautz}},
  \bibinfo{journal}{New Journal of Physics} \textbf{\bibinfo{volume}{10}},
  \bibinfo{pages}{053012} (\bibinfo{year}{2008}), ISSN
  \bibinfo{issn}{1367-2630},
  
\urlprefix\url{
http://stacks.iop.org/1367-2630/10/i=5/a=053012?key=crossref.6dda1de3a97febfeb32
d4862b2384e05}.

\bibitem[{\citenamefont{Pivetta et~al.}(2007)\citenamefont{Pivetta, Ternes,
  Patthey, and Schneider}}]{Pivetta2007}
\bibinfo{author}{\bibfnamefont{M.}~\bibnamefont{Pivetta}},
  \bibinfo{author}{\bibfnamefont{M.}~\bibnamefont{Ternes}},
  \bibinfo{author}{\bibfnamefont{F.}~\bibnamefont{Patthey}}, \bibnamefont{and}
  \bibinfo{author}{\bibfnamefont{W.-D.} \bibnamefont{Schneider}},
  \bibinfo{journal}{Physical Review Letters} \textbf{\bibinfo{volume}{99}},
  \bibinfo{pages}{126104} (\bibinfo{year}{2007}), ISSN
  \bibinfo{issn}{0031-9007},
  \urlprefix\url{http://link.aps.org/doi/10.1103/PhysRevLett.99.126104}.

\bibitem[{\citenamefont{Esat et~al.}(2015)\citenamefont{Esat, Deilmann,
  Lechtenberg, Wagner, Kr{\"{u}}ger, Temirov, Anders, Rohlfing, and
  Tautz}}]{Esat2015}
\bibinfo{author}{\bibfnamefont{T.}~\bibnamefont{Esat}},
  \bibinfo{author}{\bibfnamefont{T.}~\bibnamefont{Deilmann}},
  \bibinfo{author}{\bibfnamefont{B.}~\bibnamefont{Lechtenberg}},
  \bibinfo{author}{\bibfnamefont{C.}~\bibnamefont{Wagner}},
  \bibinfo{author}{\bibfnamefont{P.}~\bibnamefont{Kr{\"{u}}ger}},
  \bibinfo{author}{\bibfnamefont{R.}~\bibnamefont{Temirov}},
  \bibinfo{author}{\bibfnamefont{F.~B.} \bibnamefont{Anders}},
  \bibinfo{author}{\bibfnamefont{M.}~\bibnamefont{Rohlfing}}, \bibnamefont{and}
  \bibinfo{author}{\bibfnamefont{F.~S.} \bibnamefont{Tautz}},
  \bibinfo{journal}{Physical Review B} \textbf{\bibinfo{volume}{91}},
  \bibinfo{pages}{144415} (\bibinfo{year}{2015}), ISSN
  \bibinfo{issn}{1098-0121},
  \urlprefix\url{http://link.aps.org/doi/10.1103/PhysRevB.91.144415}.

\bibitem[{\citenamefont{Sader and Jarvis}(2004)}]{Sader2004}
\bibinfo{author}{\bibfnamefont{J.~E.} \bibnamefont{Sader}} \bibnamefont{and}
  \bibinfo{author}{\bibfnamefont{S.~P.} \bibnamefont{Jarvis}},
  \bibinfo{journal}{Applied Physics Letters} \textbf{\bibinfo{volume}{84}},
  \bibinfo{pages}{1801} (\bibinfo{year}{2004}), ISSN \bibinfo{issn}{00036951},
  
\urlprefix\url{
http://scitation.aip.org/content/aip/journal/apl/84/10/10.1063/1.1667267}.

\bibitem[{\citenamefont{Sader and Sugimoto}(2010)}]{Sader2010}
\bibinfo{author}{\bibfnamefont{J.~E.} \bibnamefont{Sader}} \bibnamefont{and}
  \bibinfo{author}{\bibfnamefont{Y.}~\bibnamefont{Sugimoto}},
  \bibinfo{journal}{Applied Physics Letters} \textbf{\bibinfo{volume}{97}},
  \bibinfo{pages}{043502} (\bibinfo{year}{2010}), ISSN
  \bibinfo{issn}{00036951},
  
\urlprefix\url{
http://scitation.aip.org/content/aip/journal/apl/97/4/10.1063/1.3464165}.

\bibitem[{\citenamefont{Hapala et~al.}(2014)\citenamefont{Hapala, Kichin,
  Wagner, Tautz, Temirov, and Jel{\'{i}}nek}}]{Hapala2014}
\bibinfo{author}{\bibfnamefont{P.}~\bibnamefont{Hapala}},
  \bibinfo{author}{\bibfnamefont{G.}~\bibnamefont{Kichin}},
  \bibinfo{author}{\bibfnamefont{C.}~\bibnamefont{Wagner}},
  \bibinfo{author}{\bibfnamefont{F.~S.} \bibnamefont{Tautz}},
  \bibinfo{author}{\bibfnamefont{R.}~\bibnamefont{Temirov}}, \bibnamefont{and}
  \bibinfo{author}{\bibfnamefont{P.}~\bibnamefont{Jel{\'{i}}nek}},
  \bibinfo{journal}{Physical Review B} \textbf{\bibinfo{volume}{90}},
  \bibinfo{pages}{085421} (\bibinfo{year}{2014}), ISSN
  \bibinfo{issn}{1098-0121},
  \urlprefix\url{http://link.aps.org/doi/10.1103/PhysRevB.90.085421}.

\bibitem[{\citenamefont{Bl{\"{o}}chl}(1994)}]{Blochl1994}
\bibinfo{author}{\bibfnamefont{P.~E.} \bibnamefont{Bl{\"{o}}chl}},
  \bibinfo{journal}{Phys. Rev. B} \textbf{\bibinfo{volume}{50}},
  \bibinfo{pages}{17953} (\bibinfo{year}{1994}), ISSN
  \bibinfo{issn}{0163-1829},
  \urlprefix\url{http://prb.aps.org/abstract/PRB/v50/i24/p17953{\_}1
  http://link.aps.org/doi/10.1103/PhysRevB.50.17953}.

\bibitem[{\citenamefont{Kresse and Furthm{\"{u}}ller}(1996)}]{Kresse1996}
\bibinfo{author}{\bibfnamefont{G.}~\bibnamefont{Kresse}} \bibnamefont{and}
  \bibinfo{author}{\bibfnamefont{J.}~\bibnamefont{Furthm{\"{u}}ller}},
  \bibinfo{journal}{Phys. Rev. B} \textbf{\bibinfo{volume}{54}},
  \bibinfo{pages}{11169} (\bibinfo{year}{1996}), ISSN
  \bibinfo{issn}{0163-1829},
  \urlprefix\url{http://link.aps.org/doi/10.1103/PhysRevB.54.11169}.

\bibitem[{\citenamefont{Kresse and Hafner}(1993)}]{Kresse1993}
\bibinfo{author}{\bibfnamefont{G.}~\bibnamefont{Kresse}} \bibnamefont{and}
  \bibinfo{author}{\bibfnamefont{J.}~\bibnamefont{Hafner}},
  \bibinfo{journal}{Phys. Rev. B} \textbf{\bibinfo{volume}{47}},
  \bibinfo{pages}{558} (\bibinfo{year}{1993}), ISSN \bibinfo{issn}{0163-1829},
  \urlprefix\url{http://link.aps.org/doi/10.1103/PhysRevB.47.558}.

\bibitem[{\citenamefont{Perdew et~al.}(1996)\citenamefont{Perdew, Burke, and
  Ernzerhof}}]{Perdew1996}
\bibinfo{author}{\bibfnamefont{J.~P.} \bibnamefont{Perdew}},
  \bibinfo{author}{\bibfnamefont{K.}~\bibnamefont{Burke}}, \bibnamefont{and}
  \bibinfo{author}{\bibfnamefont{M.}~\bibnamefont{Ernzerhof}},
  \bibinfo{journal}{Phys. Rev. Lett.} \textbf{\bibinfo{volume}{77}},
  \bibinfo{pages}{3865} (\bibinfo{year}{1996}), ISSN \bibinfo{issn}{1079-7114},
  \urlprefix\url{http://www.ncbi.nlm.nih.gov/pubmed/10062328}.

\bibitem[{\citenamefont{Dudarev et~al.}(1998)\citenamefont{Dudarev, Savrasov,
  Humphreys, and Sutton}}]{Dudarev1998}
\bibinfo{author}{\bibfnamefont{S.~L.} \bibnamefont{Dudarev}},
  \bibinfo{author}{\bibfnamefont{S.~Y.} \bibnamefont{Savrasov}},
  \bibinfo{author}{\bibfnamefont{C.~J.} \bibnamefont{Humphreys}},
  \bibnamefont{and} \bibinfo{author}{\bibfnamefont{A.~P.}
  \bibnamefont{Sutton}}, \bibinfo{journal}{Physical Review B}
  \textbf{\bibinfo{volume}{57}}, \bibinfo{pages}{1505} (\bibinfo{year}{1998}),
  ISSN \bibinfo{issn}{1098-0121},
  \urlprefix\url{http://link.aps.org/doi/10.1103/PhysRevB.57.1505}.

\end{thebibliography}

	\onecolumngrid
	\appendix
	\newpage
	
\section*{Supplemental Information}
\setlength{\belowcaptionskip}{-5mm}

	\setcounter{figure}{0}
        \renewcommand{\thefigure}{S\arabic{figure}}
        
	\begin{figure*}[ht!]
		\center{\includegraphics{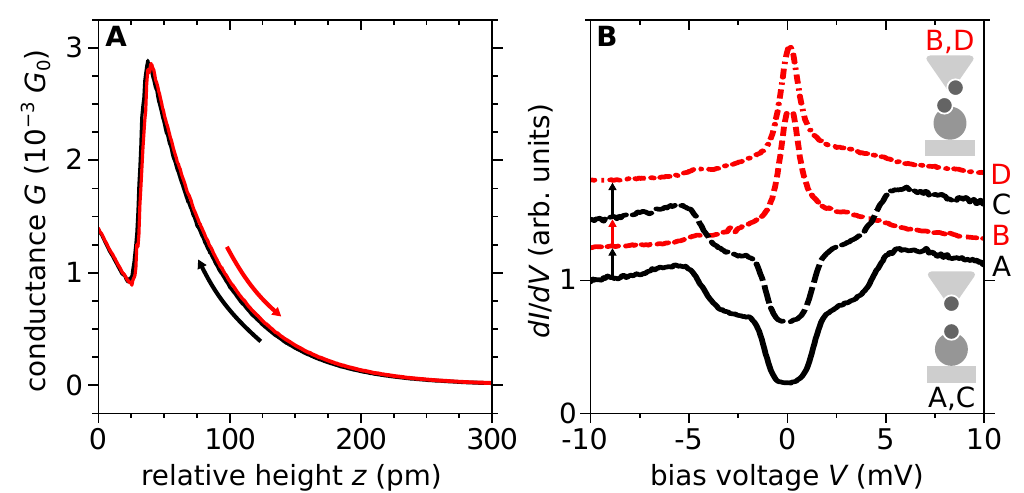}}
		\caption{\textbf{Reversibility of the Switching Process.} (A) Forward (black) and backward (red) sweeps of a conductance-distance measurement over the \chem{CoH} complex shown in Figure~\ref{fig:03} of the main text without tuning fork oscillation with $\var{V} = -10~\unit{mV}$. Approach and retract curves were taken at $5-10~\unit{pm/s}$. The switch between both spin states occurs at approximately the same \var{z}-height $(\var{\Delta z} < 25~\unit{pm})$, thus we do not observe a significant hysteresis. (B) Bias spectroscopy over the same switching complex at high ($\var{G} \sim 13\times10^{-4}~\var{G_0}$) and low conductance ($\var{G} \sim 6.5\times10^{-4}~\var{G_0}$).}
		\label{fig:S01}
	\end{figure*}
	
	\begin{figure*}[ht!]
		\center{\includegraphics{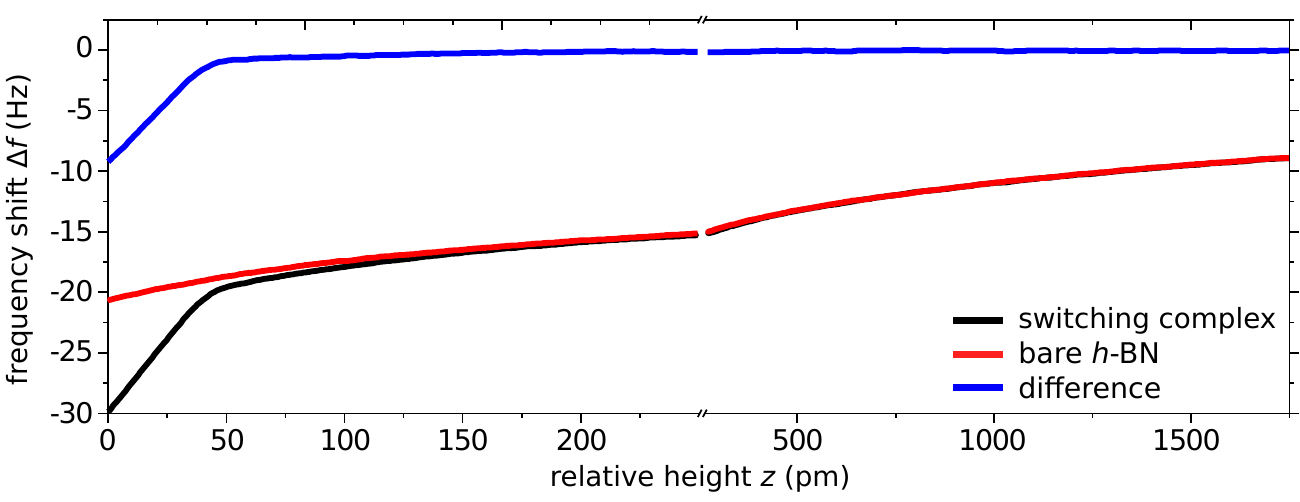}}
		\caption{\textbf{Long-Range Background Subtraction.} Raw Data of the original frequency shift curves that were taken over the switching complex in Figure~\ref{fig:03} (black) and a reference \chem{\hBN} background (red). Both curves were then substracted to account for only the short range interactions (blue).}
		\label{fig:S02}
	\end{figure*}
	
	\begin{figure*}[ht!]
		\center{\includegraphics{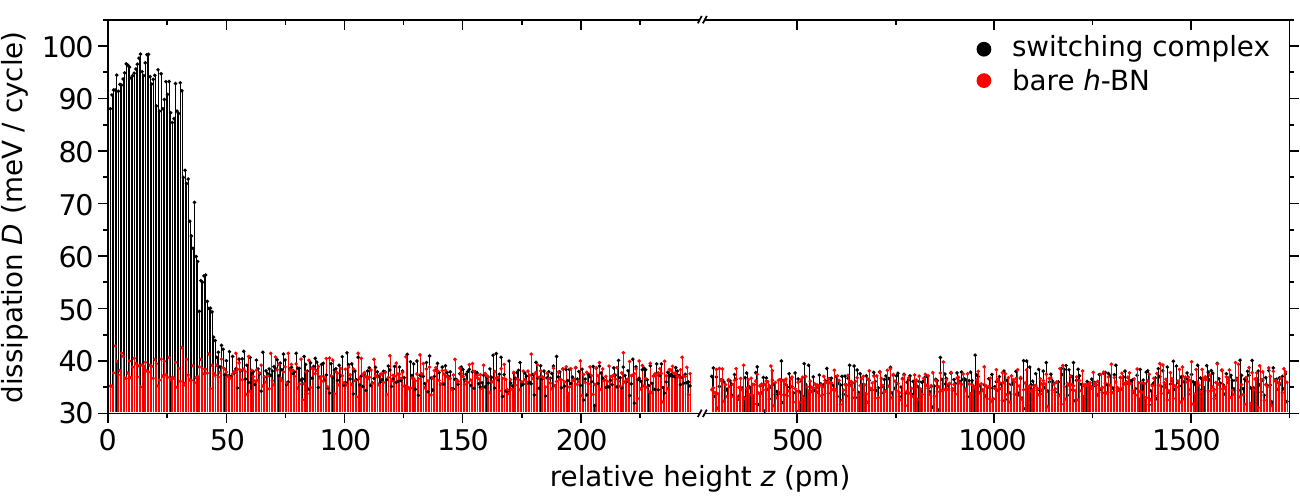}}
		\caption{\textbf{Measured Dissipation Across the Spin Transition.} Dissipation loss from the recorded excitation voltage of the frequency shift and current curves in Figure~\ref{fig:03} for the $\var{S} = 1$ to $\var{S} = \half$ switch. To find the power dissipation (Figure~\ref{fig:S02}) we calculate the energy stored in the mechanical motion of the tuning fork ($\var{E} = \frac{1}{2} \var{k_0} \var{A}^2$, \var{k_0} is the spring constant, \var{A} -- the oscillation amplitude) and find the intrinsic energy loss per oscillation cycle $\var{D} = (\pi \var{k_0} \var{A}^2)/\var{Q}$, where \var{Q} is the Q-factor of the tuning fork. For the dataset in Figure~\ref{fig:03} the parameters: $\var{Q} \approx 10000$, $\var{k_0} = 1800~\unit{N/m}$ and $\var{A} = 100~\unit{pm}$, lead to a power dissipation per cycle of around $\var{D} \approx 35~\unit{meV}$. This value increases by $\approx 55~\unit{meV}$ when the spin transition occurs at $\var{z} \approx 50~\unit{pm}$.}
		\label{fig:S03}
	\end{figure*}
	
	\begin{figure*}[ht!]
		\center{\includegraphics{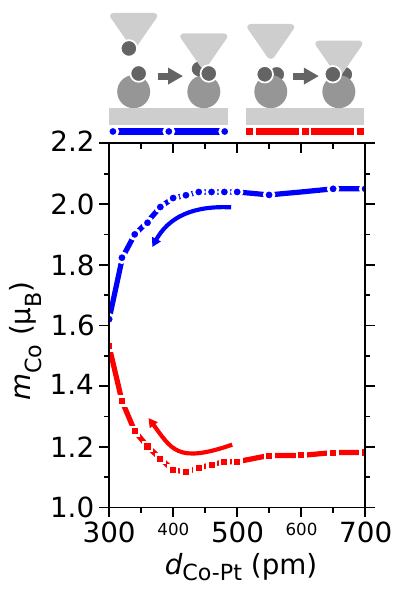}}
		\caption{\textbf{Calculated Magnetic Moment as a Function of Co--Pt Separation.} Evolution of the local magnetic moment of \chem{Co} along the diabatic tip approach curves in Figure~\ref{fig:03}D. In the case of a hydrogen-functionalized tip approaching a \chem{Co}H complex (blue line and circles), the magnetic moment is reduced as the hydrogen on the tip approaches \chem{Co} by forming a (partial) chemical bond. If a bare tip approaches a \chem{CoH_2} complex (red line, rectangles), at small junction widths the tip forms a chemical bond with both hydrogen atoms partially claiming their valency and reducing the bonding to \chem{Co}, resulting in a partial restoration of the magnetic moment.}
		\label{fig:S04}
	\end{figure*}
	
	\begin{figure*}[ht!]
		\center{\includegraphics{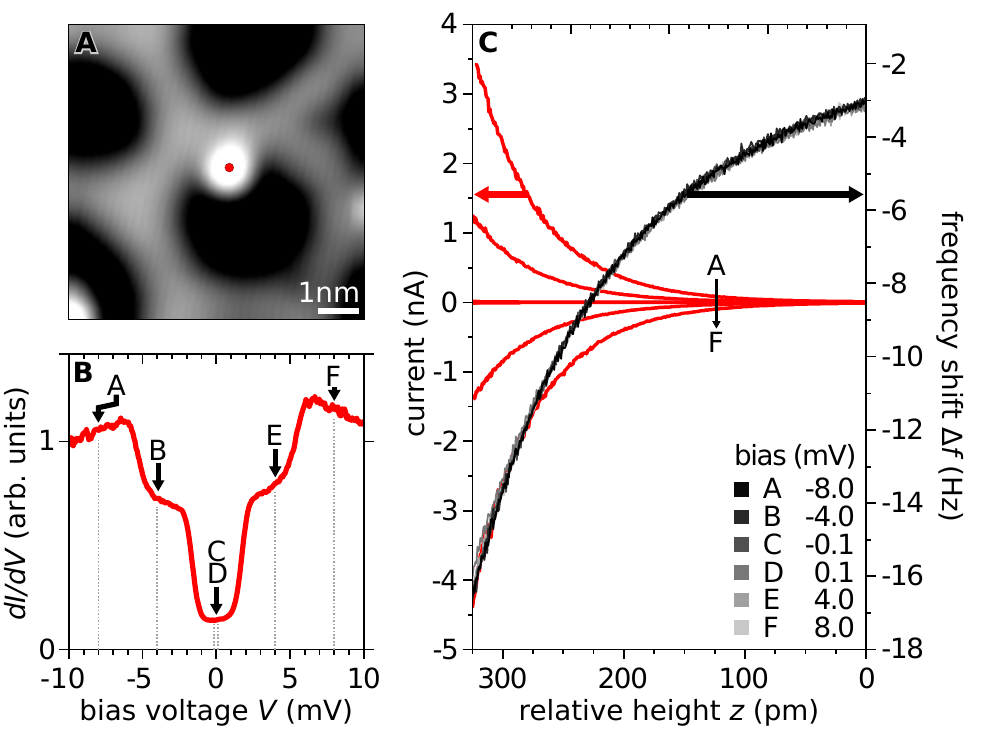}}
		\caption{\textbf{Bias and Polarity Dependent $\var{\Delta f(z)}$ Curves.} (A) Constant current STM image ($\var{V}=-15~\unit{mV}$, $\var{I} = 20~\unit{pA}$, $\var{G} = 1.72 \times 10^{-5}~\var{G_0}$) of a \chem{CoH} complex on the \chem{\hBN/Rh(111)} moir\'e obtained with a non-functionalized tip. (B) Local spectroscopy obtained on the \chem{CoH} with the tip positioned at the center of the \chem{CoH} complex (Figure~\ref{fig:S05}A, red dot) using a conductance setpoint $\var{G} = 4.30 \times 10^{-4}~\var{G_0}$. (C) Simultaneous frequency shift-distance, $\var{\Delta f(z)}$ (A--F, black--light gray), and conductance--distance, \var{G(z)} (red), measurements at different bias voltages. The tip was stabilized at $\var{V}=-15~\unit{mV}$, $\var{I} = 20~\unit{pA}$ before turning off the feedback loop and changing the bias for the different approach curves. The change in bias and polarity strongly influences the tunneling current (red curves), but has almost no effect on the frequency shift signal.}
		\label{fig:S05}
	\end{figure*}
	
	\begin{figure*}[ht!]
		\center{\includegraphics{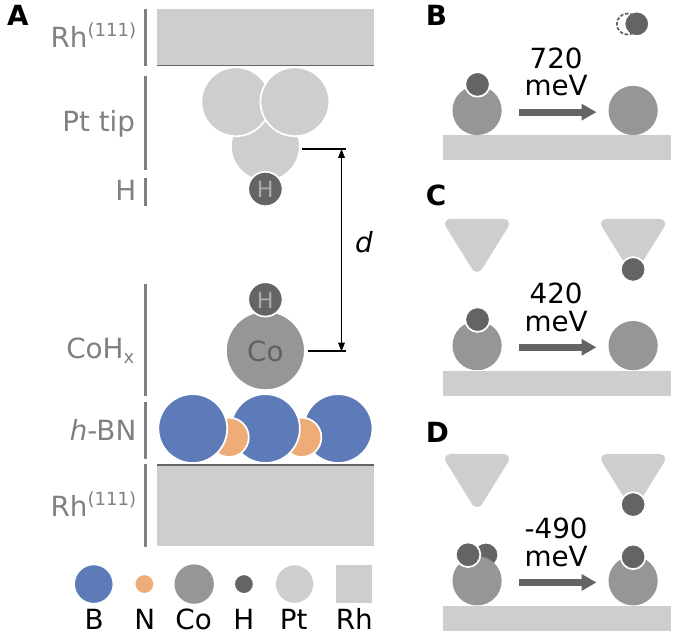}}
		\caption{\textbf{Schematic Drawing of the Simulated Junction Geometry.} (A) Schematic representation of the junction geometry used in first principles calculations: A 4-atomic functionalized \chem{Pt} tip connected to a \chem{Rh} lead suspended over a \chem{CoH} complex residing on a \chem{\hBN/Rh(111)} surface. For more details see the methods section. (B-D) Selected adsorption and transfer energies with schematic representations of the relevant processes. The adsorption energy of a single \chem{H} on \chem{Co/\hBN/Rh(111)} from molecular hydrogen gas phase, calculated as $\var{E}({\chem{H}}) = \var{E}({\chem{CoH}}) - \var{E}({\chem{Co}}) - \var{E}({\chem{H_2}})/2$, is found to be $720~\unit{meV}$ in our calculations. The transfer of an \chem{H} atom from the \chem{CoH} complex to the \chem{Pt} tip raises the total energy of the system by $420~\unit{meV}$, while the affinity of the second hydrogen in a \chem{CoH_2} complex is found to be $490~\unit{meV}$ lower than its affinity to the tip, which makes the spin-manipulation of a \chem{CoH} complex with a functionalized STM tip a reversible process.}
		\label{fig:S06}
	\end{figure*}
	
\end{document}